\documentstyle[12pt,epsf]{article}
\input{psfig}
\newlength{\dinwidth}
\newlength{\dinmargin}
\begin{document}

\def\g{\gamma}
\def\xg{$ x_\g$}
\def\xgo{$ x_\g^{OBS}$}
\def\xgom{$ x_\g^{cal}$}
\def\xglo{$ x_\g^{LO}$}
\def\xp{$ x_p$}
\def\xpo{$ x_p^{OBS}$}
\def\xplo{$ x_p^{LO}$}
\def\pT{$p_T$ }
\def\pZ{p_{\it z}}
\def\Z{{\it z}}
\def\ETC{E_T^{cone}}
\def\ra{\rightarrow}
\def\LE{$L_e^{CAL}$}
\def\LG{$L_{\gamma}^{CAL}$}
\def\ETAJ{\eta^{jet}}
\def\ETJ{E_T^{jet}}
\def\ETAP{\eta^{parton}}
\def\ETP{E_T^{parton}}
\def\ETJM{E_T^{cal}}
\def\DETA{|\Delta\eta|}
\def\ETAM{\eta^{cal}}
\def\PHIM{\phi^{cal}}
\def\EEP{E^\prime_{e}}
\def\TEP{\theta^\prime_{e}}
\newcommand{\gsim}{\buildrel{>}\over{\sim}}
\def\3{\ss}
\vspace{3cm}

\title {
{\bf Dijet Cross Sections in Photoproduction at HERA}
\\
\author{\rm ZEUS Collaboration \\}
}
\date{ }
\maketitle

\setcounter{page}{0}

\vspace{5cm}

\begin{abstract}
Dijet production by almost real photons has been studied at HERA with the
ZEUS detector. Jets have been identified using the cone algorithm.  A cut
on \xgo, the fraction of the photon energy participating in the production
of the two jets of highest transverse energy, is used to define cross
sections sensitive to the parton distributions in the proton and in the
photon. The dependence of the dijet cross sections on pseudorapidity has
been measured for \xgo $\ge 0.75$ and \xgo $< 0.75$. The former is
sensitive to the gluon momentum density in the proton.  The latter is
sensitive to the gluon in the photon.  The cross sections are corrected
for detector acceptance and compared to leading order QCD calculations.
\end{abstract}

\vspace{-19cm}
\begin{flushleft}
\tt DESY 95-033 \\
February 1995 \\
\end{flushleft}

\setcounter{page}{0}
\thispagestyle{empty}

\newpage

\textwidth15.5cm

\begin{footnotesize}
\renewcommand{\thepage}{\Roman{page}}

M.~Derrick, D.~Krakauer, S.~Magill, D.~Mikunas, B.~Musgrave,
J.~Repond, R.~Stanek, R.L.~Talaga, H.~Zhang \\
{\it Argonne National Laboratory, Argonne, IL, USA}~$^{p}$\\[6pt]
R.~Ayad$^1$, G.~Bari, M.~Basile,
L.~Bellagamba, D.~Boscherini, A.~Bruni, G.~Bruni, P.~Bruni, G.~Cara
Romeo, G.~Castellini$^{2}$, M.~Chiarini,
L.~Cifarelli$^{3}$, F.~Cindolo, A.~Contin, M.~Corradi, I.~Gialas, \\
P.~Giusti, G.~Iacobucci, G.~Laurenti, G.~Levi, A.~Margotti,
T.~Massam, R.~Nania, C.~Nemoz, \\
F.~Palmonari, A.~Polini, G.~Sartorelli, R.~Timellini, Y.~Zamora
Garcia$^{1}$,
A.~Zichichi \\
{\it University and INFN Bologna, Bologna, Italy}~$^{f}$
% ...
\\[6pt]
\clearpage

\hspace*{1mm}
$^{ 1}$ supported by Worldlab, Lausanne, Switzerland \\
\hspace*{1mm}
$^{ 2}$ also at IROE Florence, Italy  \\
\hspace*{1mm}
$^{ 3}$ now at Univ. of Salerno and INFN Napoli, Italy  \\
\hspace*{1mm}
% ...
$^{13}$ now with OPAL Collaboration, Faculty of Physics at Univ. of
        Freiburg \\
% ...

\begin{tabular}{lp{15cm}}
$^{a}$ &supported by the Natural Sciences and Engineering Research
         Council of Canada (NSERC) \\
$^{b}$ &supported by the FCAR of Qu\'ebec, Canada\\
$^{c}$ &supported by the German Federal Ministry for Research and
         Technology (BMFT)\\
$^{d}$ &supported by the MINERVA Gesellschaft f\"ur Forschung GmbH,
         and by the Israel Academy of Science \\
$^{e}$ &supported by the German Israeli Foundation, and
         by the Israel Academy of Science \\
$^{f}$ &supported by the Italian National Institute for Nuclear Physics
         (INFN) \\
$^{g}$ &supported by the Japanese Ministry of Education, Science and
         Culture (the Monbusho)
         and its grants for Scientific Research\\
$^{h}$ &supported by the Korean Ministry of Education and Korea Science
         and Engineering Foundation \\
$^{i}$ &supported by the Netherlands Foundation for Research on Matter
         (FOM)\\
$^{j}$ &supported by the Polish State Committee for Scientific Research
         (grant No. SPB/P3/202/93) and the Foundation for Polish-
         German Collaboration (proj. No. 506/92) \\
$^{k}$ &supported by the Polish State Committee for Scientific
         Research (grant No. PB 861/2/91 and No. 2 2372 9102,
         grant No. PB 2 2376 9102 and No. PB 2 0092 9101) \\
$^{l}$ &partially supported by the German Federal Ministry for
         Research and Technology (BMFT) \\
$^{m}$ &supported by the German Federal Ministry for Research and
         Technology (BMFT), the Volkswagen Foundation, and the Deutsche
         Forschungsgemeinschaft \\
$^{n}$ &supported by the Spanish Ministry of Education and Science
         through funds provided by CICYT \\
$^{o}$ &supported by the Particle Physics and Astronomy Research
        Council \\
$^{p}$ &supported by the US Department of Energy \\
$^{q}$ &supported by the US National Science Foundation
\end{tabular}

\textwidth17.0cm
\newpage
\setcounter{page}{1}
\renewcommand{\thepage}{\arabic{page}}

\end{footnotesize}

\section{Introduction}

\begin{figure}
\setlength{\unitlength}{1mm}
\epsfysize=400pt
\epsfbox[50 200 450 600]{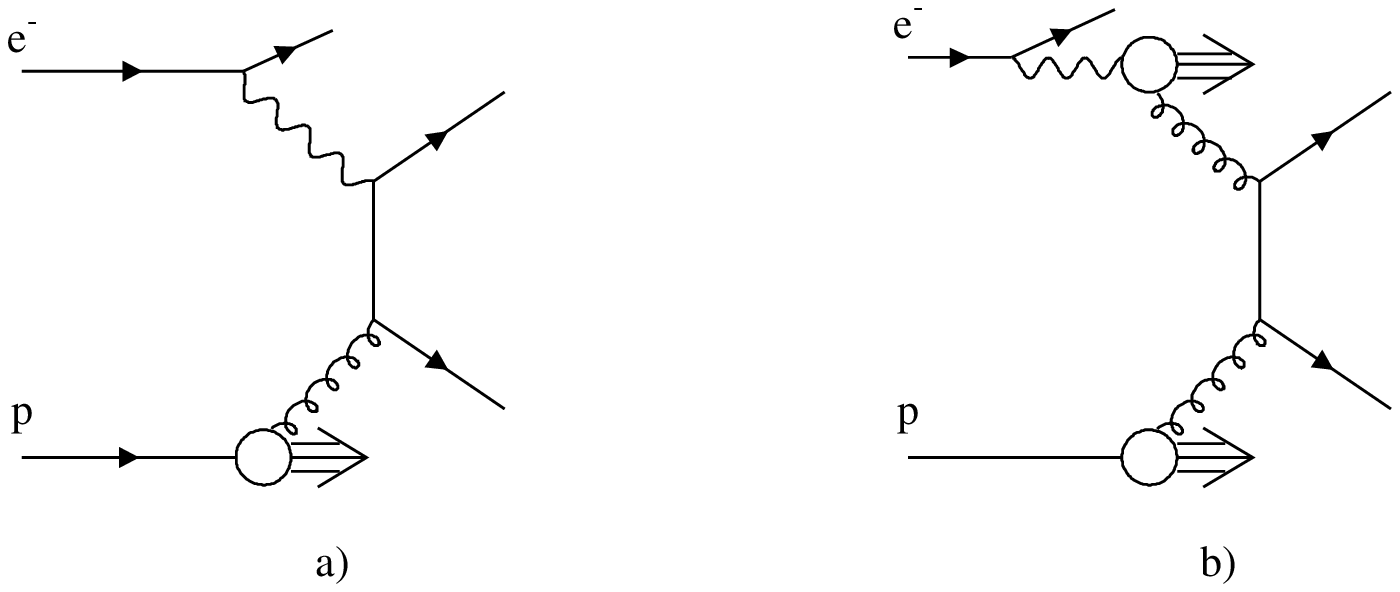}
\caption{\label{f:diag}
        { Examples of leading order diagrams for a) direct and b) resolved
photoproduction.}}
\end{figure}

At the high energies available at HERA, interactions between almost real
photons
(of virtuality $Q^2 \approx 0$) and protons
produce jets of high transverse energy~\cite{H11,Z1,H12,Z2,Z3}.
The presence of a `hard' energy scale means that perturbative QCD
calculations of event properties can be confronted with experiment.
At leading order (LO)
%${\cal O}(\alpha\alpha_s$)),
two processes are
responsible for jet production. The photon may interact directly with
a parton in the proton (Fig.~\ref{f:diag}a), or it may first fluctuate into
an hadronic state (Fig.~\ref{f:diag}b). In the first case, known as the
direct contribution, the full energy of the
photon participates in the interaction with a parton in the
proton and the
fraction of the photon momentum ($x_\gamma$)
participating in the hard process is equal to one.
The final state of the direct process consists of two jets,
the proton remnant and the scattered electron.
In the second case, known as the resolved contribution, the photon
acts as a
source of partons which then scatter off partons in the proton
and the
fraction of the photon momentum
participating in the hard process is less than one.
The final state in this case includes
a photon remnant, continuing in the original
photon direction, in addition to two jets, the proton remnant and the
scattered electron.
At higher orders this simple distinction between direct and resolved is no
longer
precisely defined.

With a cut on jet transverse energy of $E_T^{jet} > 6$~GeV, direct photon
events probe the parton distributions in the proton down
to \xp~ $\approx 2 \times 10^{-3}$, where \xp\ is the fraction of the
proton's momentum entering into the hard process.
This process is directly sensitive to the gluon distribution
in the proton, and complements indirect extractions
in the same \xp\ range~\cite{F2gluon}
using the measurement of $F_2$ in deep inelastic scattering (DIS).
The \xp\ values sampled by the resolved photon contribution are typically
higher
than those of the direct contribution,
lying in a region where the proton parton distributions are
constrained by DIS data.
Resolved photon processes are directly sensitive to the photon
parton distributions, particularly the gluon distribution in the photon,
down \xg~ $\approx 0.06$.
This is not the case in the measurement of $F_2^\gamma$
in $\gamma\gamma$ interactions
at $e^+e^-$
colliders, where highly virtual photons are used to probe almost real
photons~\cite{alex}.
In collisions between two almost real photons, however,
the measurement of
jet cross sections
has recently shown potential to constrain the gluon distribution in
the photon~\cite{XLAC3,TOPAZ}.

In this paper we separate the direct and resolved
photon contributions to jet production, and present
dijet differential $ep$ cross sections
which are sensitive to the gluon distributions in the
proton and photon respectively.
In describing the cross sections to be
measured, particular attention is paid to the way in which direct and
resolved processes are defined.
% We then describe the experimental setup and event
% selection.
Differential $ep$ cross sections are presented as a function of jet
pseudorapidity for direct and resolved photon processes and compared to
available LO QCD calculations.

\section{Definition of cross sections}

We measure the cross section for dijet photoproduction,
$ep~\rightarrow~e\gamma~p~\rightarrow~e~X$, for
events
in which $X$ contains
at least two jets
of $E_T^{jet} > 6$~GeV.
In this experiment, photoproduction events are
defined by demanding that the
electron is scattered at small angles and does not emerge from the beam pipe.
This requirement corresponds approximately to a cut of
$Q^2 < 4$~GeV$^2$, giving a
median $Q^2$ of
$\sim 10^{-3}$ GeV$^2$~\cite{Z2}.
% 0.001~GeV$^2$~\cite{Z2}.
The cross section measured is
$d\sigma/d\bar{\eta}$, where
$\bar{\eta} = \frac{1}{2}(\eta_1 + \eta_2)$ is the average
pseudorapidity\footnote{The $z$ axis is defined to
lie along the proton
direction, and $\eta = - $ln$(\tan\frac{\theta}{2}$) where $\theta$ is
the angle between the jet and the $z$ axis.}
of the two jets of highest
transverse energy,
with the requirement that
$\DETA = |\eta_1 - \eta_2| < 0.5$.
The cross section is measured for $0.2 < y < 0.8$, where $y = E_\gamma/E_e$
is the fraction of the initial electron energy
($E_e$) carried by the almost real photon with energy $E_\gamma$.
This $y$ interval corresponds to $\gamma p$ centre-of-mass energies
($W_{\gamma p}$) in the range 132~GeV $< W_{\gamma p} < 265$~GeV.

The possibility of experimentally separating
samples of direct and resolved photon
events was demonstrated in~\cite{Z2}. However, as the simple
definition of resolved and direct photoproduction is only unambiguous at
leading order, it is important to find a definition which is both
calculable to all orders and measurable.

For two-to-two parton scattering in LO QCD,
energy and momentum conservation give the fraction of the photon energy
involved in the hard scatter as

\begin{equation}
x_\g^{LO} = \frac{ \sum_{partons}E_T^{parton}e^{-\eta^{parton}}}{2yE_e},
\label{xgloeq}
\end{equation}

\noindent where $yE_e$ is the initial photon energy and the sum is
over the two final state partons. For direct photon events, $x_\g^{LO} = 1$.
Since it is not possible to measure partons, we define an observable in
terms of jets which is analogous to \xglo. This observable, called
\xgo, is the fraction of the photon's momentum participating in the
production of the two highest $\ETJ$ jets. The explicit definition is,

\begin{equation}
x_\g^{OBS} = \frac{ \sum_{jets}E_T^{jet}e^{-\eta^{jet}}}{2yE_e},
\label{xgoeq}
\end{equation}

\noindent where now the sum runs over the two jets of highest $\ETJ$.
In the \xgo~ distribution thus obtained, the
LO direct and resolved processes populate different regions, with the
direct processes concentrated at high values of \xgo. The peak arising from
the direct contribution will not necessarily lie exactly at \xgo $= 1$ due to
higher order effects and/or hadronisation, but will still correspond to the
kinematic region where most or all of the energy of the photon is
available to probe
the proton. The relationship between the definitions of \xgo~and
\xglo~ is dependent upon the hadronisation and showering
models used to describe the final state
and the assumptions made in the calculation of the hard process,
as \xglo~ becomes ambiguous even at
next-to-leading order (NLO).
For these reasons, we will present hadronic jet cross sections and will
not use Monte Carlo models to correct back to parton kinematics.
For the purposes of this paper,
the separation between direct and resolved photoproduction is defined by a cut
on \xgo. Other model-independent definitions for separating resolved and
direct photon processes have been suggested, often using a cut on the
energy in a cone around
the photon direction~\cite{dag}. We have chosen the definition based upon
\xgo~because it depends only upon the measurement of $y$,
$\ETJ$ and the $\eta$'s of the jets, without the need to introduce further
variables.

The $\DETA$ cut ensures that for the
$\bar{\eta}$ bins in which the cross sections are measured, both jets lie
well within the acceptance of the ZEUS detector, without imposing additional
cuts at high and low $\ETAJ$.
Applying this cut has two further benefits, which can be seen
as follows. When both jets are at equal $\eta$ and equal $E_T^{jet}$,
$\sqrt{\hat{s}} = 2\ETJ$
and so the range of possible choices of scale in theoretical calculations
is reduced. More importantly, by rewriting the expression
for \xgo\ in terms of
$\DETA$ and $\bar\eta$, and assuming the jets to have equal transverse
energy, one obtains

\begin{equation}
x_\g^{OBS} = \frac{E_T^{jet}e^{-\bar{\eta}}}{yE_e} \cosh \frac{\Delta \eta}{2}.
\label{xgoeq2}
\end{equation}

A similar expression can be written for the proton,

\begin{equation}
x_p^{OBS}  = \frac{E_T^{jet}e^{\bar{\eta}}}{E_p} \cosh \frac{\Delta \eta}{2},
\label{xgoeq3}
\end{equation}

where $E_p$ is the incident proton energy.
When the jets are at equal pseudorapidities
the hyperbolic cosine term takes its minimum value of unity.
Thus the minimum available
$x$ values are probed for a given $\ETJ$ and $\bar{\eta}$, and there
is a strong correlation between $\bar{\eta}$ and \xpo\ in the direct cross
section and between $\bar{\eta}$ and $y x_\g^{OBS}$ in the resolved cross
section~\cite{jeff1}. Applying the cut $\DETA < 0.5$ brings us close
to this situation.

\section{The ZEUS detector and beam conditions}

Details of the ZEUS detector have been described elsewhere~\cite{ZEUS}.
The primary components used in this analysis are the calorimeter and the
tracking detectors. The
uranium-scintillator calorimeter \cite{CAL} covers about 99.7\% of the
total solid angle and is subdivided into electromagnetic and hadronic
sections with cell size, respectively,
of $5 \times 20$ cm$^2$ ($10 \times 20$ cm$^2$
in the rear calorimeter {\it i.e.} the electron direction),
and $20 \times 20$ cm$^2$.
The calorimeter has an equal response to electrons and hadrons within 3\%
and the energy resolution achieved in test beams, in terms
of the energy $E$ in GeV, is $\sigma/E$ =
18 \%/$\sqrt E$ for electrons and $\sigma/E$ =35 \%/$\sqrt E$ for hadrons.
The timing resolution of a calorimeter cell is better than
$\sigma_t$ =1.5/$\sqrt E$ $\oplus$ 0.5 ns.

The tracking system consists of a vertex detector (VXD)~\cite{VXD}
and a central
tracking chamber (CTD)~\cite{CTD}  enclosed in a 1.43 T solenoidal
magnetic field. The interaction vertex
is measured with a resolution along (transverse to) the beam direction
of 0.4~(0.1)~cm.

To allow a precise measurement of the luminosity
via the electron-proton Brems\-strahlung process,
electron and photon
lead-scintillator calorimeters have been installed
inside the HERA tunnel, subtending small angles from the interaction
vertex~\cite{LUMI}.
The small-angle electron calorimeter is also used to tag a subsample of
photoproduction events in the approximate range
$10^{-8}$~GeV$^2 < Q^2 < 10^{-2}$~GeV$^2$.

In 1993 HERA operated with 84 colliding bunches
of 820~GeV protons and 26.7~GeV electrons
with typical
beam currents around 10 mA and a luminosity
of $0.6 \times 10^{30}$ cm$^{-2}$ s$^{-1}$.
{}From these colliding bunches ZEUS collected
a total integrated luminosity of approximately 0.55~pb$^{-1}$.
Additional unpaired electron and proton
bunches circulated to allow monitoring of background
from beam-gas interactions.

\section{Data selection and jet finding}

ZEUS data acquisition uses a three level trigger system.
During the 1993 data taking period, we have selected events
which were triggered at the first level on
regional or transverse energy sums in the uranium calorimeter.
Events were also triggered on a coincidence of an electron
measured in
the small-angle electron calorimeter and
an energy deposit in the rear calorimeter.
These events were used to check the efficiency of the
calorimeter-only triggers used in the cross section calculation.

At the second level trigger, cuts on calorimeter timing were used to
remove events caused by interactions between the proton beam and residual
gas in the
beam pipe in front of the calorimeter~\cite{F294}.
The efficiency of the combination of the first and second level triggers
to select events in the kinematic region of our cross sections
has been determined to be greater than $98\%$.
%First level triggers caused by discharges in the
%calorimeter
%photomultiplier tubes were also removed.

At the third level trigger, tighter timing cuts were made to reject
events arising from proton beam-gas interactions.
Cosmic ray events were vetoed using information from the tracking chambers and
calorimeter. Events
with no vertex found by the central tracking chambers, or with a vertex
found at $z < -75$~cm were rejected.
The measured transverse energy outside a $10^\circ$ cone around the forward
beampipe
($\ETC$) was calculated, making use of the vertex information from the
tracking
detectors and associating the energy in a cell with the geometric centre
of that cell.
Events with $\ETC < 12$~GeV were rejected. After these triggers,
470,000 events remained.
% corresponding to
% a total integrated luminosity of approximately 0.55~pb$^{-1}$.

For the final analysis, more stringent  cuts using calorimeter
timing and tracking information are made to further reduce
the background from cosmic rays and beam-gas interactions.
Two additional cuts are made, based upon different measurements of
$y$~\cite{Z2}:
\begin{enumerate}
\item
Events with an electron candidate in the uranium calorimeter
are removed if
$\EEP$ is greater than 5~GeV, and if
the electron gives a measured $y_e = 1 - \frac{\EEP}{2 E_e}
(1-\cos{\TEP} ) < 0.7$,
where $\EEP$ and $\TEP$ are the
energy and angle of the scattered electron.
\item
A cut is made on the Jacquet-Blondel measurement of $y$,
%$y_{JB} = \sum_i E_i(1-\cos\theta_i)/2E_e$,
$y_{JB} = \sum_i (E_i - E_{zi}) /2E_e$,
where $E_{zi} = E_i \cos(\theta_i)$, and
$E_i$ and $\theta_i$ are the energy and polar angle
of the calorimeter cell.  The angle is
determined using the measured $z$-vertex of the event.
It is assumed that the scattered electron is
not seen in the uranium calorimeter and so the sum runs over all
calorimeter cells.
%  of energy and angle $E_{cell}, \theta_{cell}$.
For any remaining events for which the scattered electron did enter the
uranium calorimeter and either was not identified or gave
$y_e$ above $0.7$,
the value of $y_{JB}$ under this assumption will be
near one. Proton beam-gas events will have low values of $y_{JB}$.
To further reduce
contamination from both these sources,
we demand that $0.15 < y_{JB} < 0.7$, which is an
estimator of the actual $y$ interval of $0.2 < y < 0.8$,
as determined from studies of the
energy loss in inactive material in front of the uranium calorimeter.
\end{enumerate}

For the surviving events, jet finding is performed on all cells of the
uranium calorimeter
using a  cone algorithm~\cite{snow}. The cone radius of
the jet finding algorithm is defined such that
\newline $R = \sqrt{\Delta\eta_{cell}^2 + \Delta\phi_{cell}^2} = 1$.
The algorithm searches in pseudorapidity-azimuth ($\eta_{cell}$-$\phi_{cell}$)
space for the cone containing the highest
summed $E_T$, removes the cells in this cone and continues the search for
the next
highest $E_T$ cone.
If the summed transverse energy of the calorimeter cells within a cone
($\ETJM$) is greater than 5~GeV, and if the $E_T$ weighted
pseudorapidity of the centre of the cone lies
within $-1.125 < \ETAM < 1.875$, then the cells
inside
this cone are selected as a jet.
The $E_T^{cal}$ threshold of 5 GeV is to compensate for energy losses
from inactive material in front of the
calorimeter.  This energy loss
is corrected for in the determination of the cross sections.
This $\eta$ range is dictated by the available statistics and by the
necessity
to remain in the central region of the calorimeter.
Events are selected if they have at least two jets satisfying these cuts.

After these cuts a sample of 12,100 photoproduction events remains.
This is reduced
to a final sample of 4,000 events after a cut on the value of $\Delta\eta$
reconstructed from the two jets of highest transverse energy found in the
calorimeter ($|\Delta\eta^{cal}| < 0.5$).
The backgrounds from cosmic rays and beam-gas interactions
are 0.6\% and 1.0\%,
respectively, as
determined from the number of surviving events originating from empty bunch
crossings and unpaired proton bunches.
The contamination from
events with a scattered electron in the uranium calorimeter
is found to be around 2.5\% from studies using DIS Monte Carlo events.
Studies using the independently selected sample for which the scattered
electron is seen in the small angle
electron calorimeter and for which lower energy thresholds in the uranium
calorimeter are
applied, show that the combined efficiency of the first level trigger
and the $\ETC$ cut is
greater than $90\%$ for events with two jets of $E_T^{cal} >$ 5 GeV.

\section{Resolution of kinematic variables}

In order to estimate how well jets and energies are reconstructed in the ZEUS
detector,
and to study the efficiency of the data selection cuts,
we have used the HERWIG 5.7~\cite{HRW} and PYTHIA~5.6~\cite{PYT} event
generators in
conjunction with a detailed simulation of the ZEUS detector and triggers.
The simulated integrated luminosity used is greater than
that of the data.
As shown in~\cite{Z2}, these generators are able to
provide a reasonable
description of the data with
the minimum $\hat{p}_T$ of a hard scatter
set to 2.5~GeV and this value was used in the current implementations.
The parton
distribution sets used were
MRSD$_-$~\cite{MRS} for the proton
and GRV LO for the photon~\cite{GRV}.
For each generator (HERWIG and PYTHIA), samples of direct and
resolved photoproduction events were
combined according to the generated cross sections.
The Monte Carlo simulations are only used to correct
for detector acceptance and
smearing.

The experimental shifts and resolutions predicted by the Monte Carlo
simulations were
obtained by comparing the true variables with the reconstructed values in the
simulated detector. The measured value of \xgo\ is called \xgom\ and is
calculated by inserting $E_T^{cal}$, $\eta^{cal}$, and $y_{JB}$ into
Equation~(\ref{xgoeq}).
For the jet variables (including \xgo), the true
variables are calculated by performing jet finding on the
final state particles
%hadronic final state
%final state hadrons
generated by the HERWIG
or PYTHIA  program before detector simulation.
The reconstructed value of $y$ ($y_{JB}$) is systematically lower than the true
value by an average of 0.08, and has a resolution of 0.06. The measurements
of $\eta$,
$\bar\eta$, $\Delta\eta$ and
\xgo\ exhibit no systematic shift and have resolutions of 0.06 (0.07),
0.05 (0.06), 0.10 (0.13) and 0.06 (0.07)
respectively, for HERWIG (PYTHIA).
The reconstructed value of $\ETJ$ is
systematically lower than the true value by 15\% (16\%) for the HERWIG (PYTHIA)
simulation with a resolution of $12 \%$.

The description of the energy response of the uranium calorimeter
in the Monte Carlo simulation has been checked for events with an electron
measured
in the small-angle electron calorimeter.
The photon energy calculated from $E_{\gamma} = E_e - \EEP$
was compared with the value reconstructed from the calorimeter
variable $y_{JB}$
($E_{\gamma}^{cal} = y_{JB}E_e$). In
addition, for jets in the central region of the calorimeter,
the measurement of the transverse energy available from the ZEUS
tracking detectors has been used to check the simulation of the
calorimeter energy response~\cite{Z3}. The transverse energy of jets
in the forward region of the calorimeter (outside the acceptance of the
ZEUS central tracking detector) has been
compared with the $\ETJM$ of these central jets. From these investigations
we conclude that the Monte Carlo description of the energy response
of the calorimeter to jets of hadrons is correct to within
a possible overestimation of the energy of 5\%.

The effects of  discrepancies between the Monte Carlo simulations
and the data on the measurement of the cross sections
are estimated by varying the selection cuts made on
the reconstructed kinematic variables,
both in the data and the simulated sample
and are included
in the
systematic errors on the final cross section values.

\section{Results and discussion}

\begin{figure}
\vspace{1.0cm}
\epsfysize=400pt
\epsfbox[50 220 450 620]{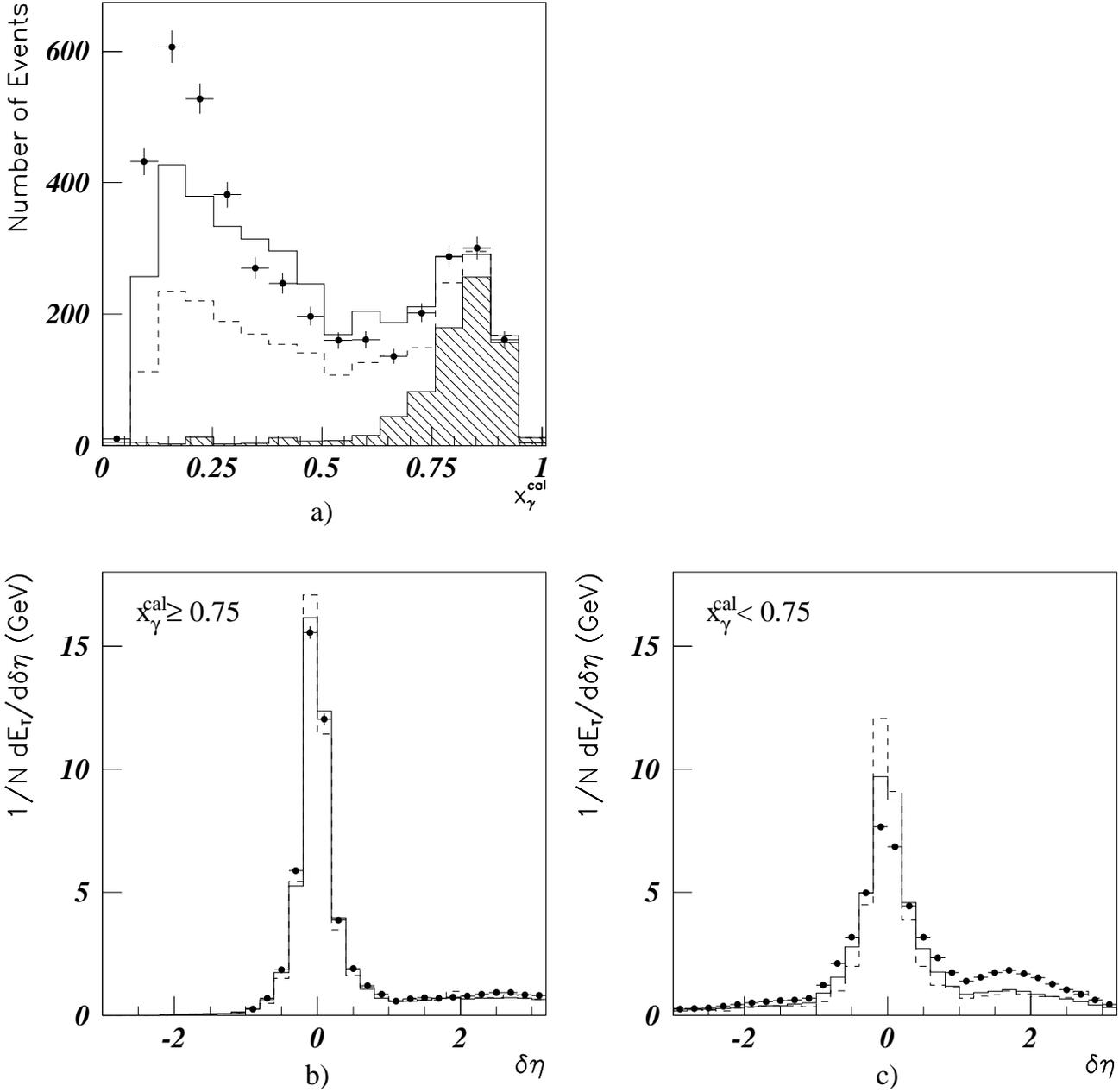}
\vspace{2.5cm}
\caption{\label{f:xg}{a) The \xgom~ distribution.
The solid circles are uncorrected ZEUS data.
The solid (dashed) line represents the distribution from the PYTHIA (HERWIG)
simulation.
The LO direct contribution to the
HERWIG distribution is shown by the shaded histogram.
The Monte Carlo curves have been normalised
to fit the direct peak in the data.
b) and c) show the uncorrected transverse energy
flow $1/N dE_T/d\delta\eta$ around the jet axis,
for cells within one
radian in $\phi$ of the jet axis, for b) direct and c) resolved events.
The solid (dashed) line represents the distribution from PYTHIA (HERWIG).}}
\end{figure}

Fig.~\ref{f:xg}a shows the distribution of \xgom~for the final data
sample entering into the cross section measurements.
There is a clear peak at high values of $x_\gamma^{cal}$
which indicates the presence of direct type processes.
The shape of this uncorrected $x_\gamma^{OBS}$ distribution and the
position of the peak
are insensitive to the problem of calorimeter energy response because
of the presence of energy terms in both the numerator and denominator
of Equation~(\ref{xgoeq}).
The distributions from PYTHIA and HERWIG
are also shown where
the Monte Carlo curves have
been normalised
to fit the direct peak in the data.
%the generated (LO) direct
%and resolved components have been fitted to the data with independently free
%normalisations as in~\cite{Z2}.
%This distribution is particularly dependent on the photon parton distribution.
Although the shape of the direct peak is reasonably reproduced,
the Monte Carlo simulations fail to describe the rise seen in
the data at low \xgom. This effect was also seen in~\cite{Z2}.
In the same Fig., the LO direct contribution from HERWIG
is shown separately, and indicates that
defining the direct process with a cut on \xgo~ of 0.75 corresponds well to the
LO definition of direct photon processes as implemented in the simulations.

Fig.~\ref{f:xg}b shows the uncorrected transverse energy flow per jet
$1/N dE_T/d\delta\eta$ around the jet axis for events with \xgom $\ge  0.75$.
For this class of events, both HERWIG and PYTHIA
reproduce the data distribution well.
The same distribution is shown in
Fig.~\ref{f:xg}c
for events with \xgom $< 0.75$. In this case both simulations
fail to describe the transverse energy flow in the forward region,
as was also observed for the jets in the inclusive jet cross sections
in~\cite{H12} and~\cite{Z3}.

\subsection{Direct photon dijet cross section}

The direct photon cross section was evaluated from our data by
applying acceptance correction factors obtained using the Monte Carlo
simulations. Events were selected by cuts
on the reconstructed kinematic quantities as
outlined in Section~4 and a cut
of \xgom$\ge 0.75$. For these cuts, we have evaluated, bin by bin, the
acceptance correction for the measurement of the cross section
$d\sigma/d\bar\eta$ as defined
by the kinematic variables described
in Section~2 with a cut of \xgo $\ge 0.75$.
The efficiency and purity are evaluated for bins of width 0.25,
with centres in the range $-0.75 < \bar{\eta} < 1.0$,
using the simulation of the ZEUS detector in conjunction with events
from the HERWIG and PYTHIA generators.

\begin{figure}
\vspace{1.0cm}
\epsfysize=400pt
\epsfbox[50 100 450 500]{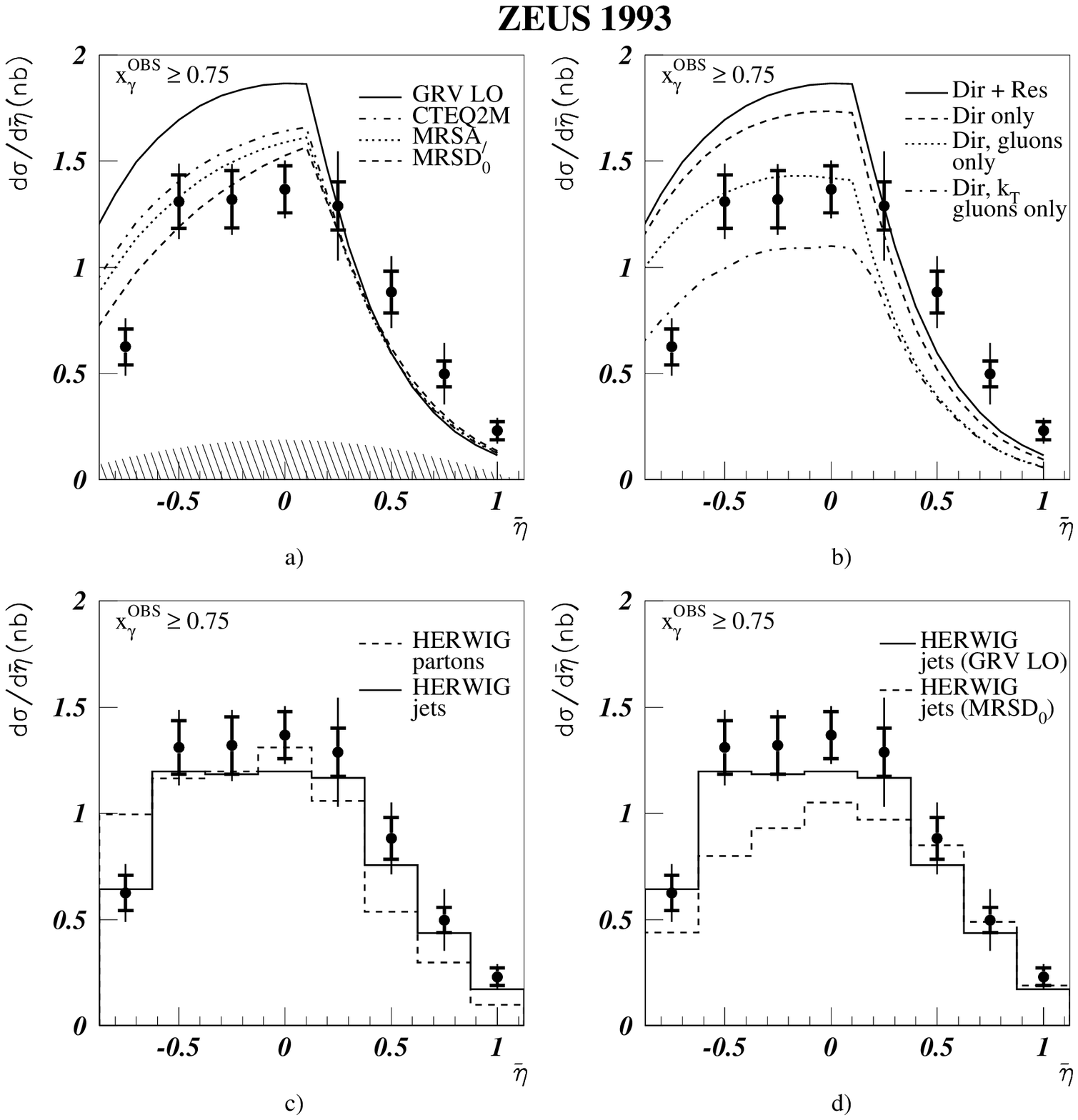}
\vspace{2.5cm}
\caption{\label{f:final2}
           { $d\sigma/d\bar{\eta}$ for $ep \rightarrow eX + \mbox{2 (or more)
jets},
\DETA < 0.5,
\ETJ > 6$~GeV,
$0.2 < y < 0.8,
Q^2 < 4$~GeV$^2$,
\xgo $\ge 0.75$.
The solid circles are corrected ZEUS data.
The inner error bars indicate the
statistical errors, the outer error
bars show the systematic uncertainty
(excluding the correlated uncertainty)
added in quadrature.
The shaded band shows the correlated uncertainty from measurement of
energy in the calorimeter and the integrated luminosity.
In a) the data are compared to LO QCD
calculations using several parton
distribution sets for the proton and the GS2 set for the photon.
In b) the data are compared to the LO QCD calculation of a) using the GRV LO
parton distribution set for the proton,
the same calculation but omitting the resolved contribution
with \xglo$ \ge 0.75$,
including only gluon induced direct photon processes and the `$k_T$
factorized' curve.
In c) the data are compared to HERWIG Monte Carlo estimates of the cross
section
using partons and final state jets.
For these HERWIG histograms
we have used the the GRV LO (LAC1) proton (photon) parton
distribution set. In d) the data are compared to HERWIG jet cross sections
using the GRV (LAC1) and MRSD$_0$ (LAC1) proton (photon) parton distribution
sets.}}
\end{figure}

The efficiency is a slowly varying function
of $\bar{\eta}$ and is around $50\%$, falling to 35\% in the lowest
$\bar{\eta}$ bin.
The purity in each bin
is approximately independent of $\bar{\eta}$ and around $60\%$ and
depends mostly on migrations across the $E_T^{jet}$ cut.
Taking the ratio of these gives an acceptance correction factor
which averages around 1.3 and rises to around 1.5 in the lowest $\bar{\eta}$
bin. This
correction accounts for all detector effects and migrations. We correct back
to the
final state particles
and no subtraction of jet pedestal energy ({\it i.e.} possible transverse
energy around the jet direction which is not associated with the hard
subprocess) is performed.
The differential cross section $d\sigma/d\bar{\eta}$ for
$x_\gamma^{OBS} \ge 0.75$
%direct photoproduction
and the kinematic range indicated in the figure caption,
is shown
in Table~\ref{t:sys2} and
in
Fig.~\ref{f:final2} where the cross section
value is plotted in the centre of the bins.
%\footnote{Numerical values for
%these cross sections can be found in a table in the preprint DESY 95-XXX.}.

\begin{table}
\centering
\begin{tabular}{|r|c|c|c|c|}  \hline
$\bar{\eta}$ & $d\sigma/d\bar{\eta}$ & Statistical & Systematic & Energy Scale
Uncertainty \\
             &         (nb)         & Error (nb)  & Error (nb) & (nb)
   \\ \hline
  -0.75      &    0.63              & 0.08        & 0.16       & 0.08
   \\
  -0.50       &    1.31              & 0.13        & 0.10       & 0.19
    \\
  -0.25      &    1.32              & 0.14        & 0.08       & 0.20
   \\
   0.00       &    1.37              & 0.11        & 0.06       & 0.24
    \\
   0.25      &    1.29              & 0.11        & 0.17       & 0.15
   \\
   0.50       &    0.88              & 0.10        & 0.17       & 0.16
    \\
   0.75      &    0.50              & 0.06        & 0.27       & 0.06
   \\
   1.00       &    0.23              & 0.04        & 0.17       & 0.05
    \\  \hline
\end{tabular}
\caption{\label{t:sys2} $d\sigma/d\bar{\eta}$ for $ep \rightarrow eX +
\mbox{2 (or more) jets},
\DETA < 0.5,
\ETJ > 6$~GeV,
$0.2 < y < 0.8$,}
$Q^2 < 4$~GeV$^2$, \xgo $\ge 0.75$.
\end{table}

The systematic uncertainties in the measurement of the differential cross
section have
been estimated by repeating the acceptance correction with both HERWIG and
PYTHIA, and by
varying the cuts made on the reconstructed quantities,
and by using different
parton distribution sets in the simulations. In addition, the bin by bin
correction procedure was checked against the results of an unfolding method
based upon Bayes' theorem~\cite{bayes}, which showed essentially
the same result.
The largest shift in each bin from these
variations was found to be similar to the statistical error on the data, and
is taken to be the total systematic error.

The systematic uncertainty arising from a possible 5\% uncertainty
%underestimation of the shift in energies measured by the calorimeter
%in the Monte Carlo simulations is highly correlated
in the mean energies measured by the calorimeter
is highly correlated
between bins,
and is therefore excluded from the systematic errors and
shown separately as a shaded band in Fig.~\ref{f:final2}a.
Also included in this shaded band is the uncertainty in the measurement of the
integrated luminosity of 3.3\%.
The shaded band represents the width of
the uncertainty around each data point.

The measured direct dijet cross section is shown
in Fig.~\ref{f:final2}a. The cross section is
around 1.3~nb at negative $\bar{\eta}$ values, and exhibits a
sharp drop near $\bar{\eta} = 0$ which
arises from the cutoff on the minimum $\ETJ$ and the cuts on $y$.
The cross section is compared to several LO QCD calculations
in which the two final state partons are
considered to be jets. The shape of the direct cross section differs
from that of the LO QCD calculations.
However, several effects
must be considered when comparing data and theory:
%influence the comparison between data and theory.
\begin{itemize}
\item
The gluon distribution of the proton in this kinematic region is not well
known. In Fig.~\ref{f:final2}a the data are compared with
LO cross section curves~\cite{jeff1} calculated using
the GRV LO, CTEQ2M~\cite{CTEQ}, MRSA~\cite{MRSA} and
MRSD$_0^\prime$ parton
distribution sets for the proton.
The MRSA and CTEQ2M parton distribution sets
are global fits to data which include HERA measurements of the structure
function $F_2$~\cite{F2PDFS}.
The contribution from the tail of the LO resolved cross
section with \xglo $\ge 0.75$ is included in the calculations
using the GS2~\cite{GS} photon parton distribution set.
This contribution to the LO cross section is small
($d\sigma/d\bar{\eta}
\approx 0.1$~nb).
This is also the case using other available photon parton distribution sets
except for the
LAC3 set~\cite{LAC}, where due to the high
gluon density at high \xglo~ the cross section becomes large enough to
describe the
whole high \xgo~ cross section without any direct component. This parton
distribution
set is disfavoured by results presented in this paper (see below)
and by other measurements~\cite{XLAC3}.
\item
QCD calculations of this cross section are only available at LO.
Some estimate of the size of higher order corrections
can be made by comparing the LO calculation using the GRV LO parton
distribution set with the other LO calculations using NLO parton distribution
sets, which
in general are around 20\% lower\footnote{The curve calculated with the GRV
NLO parton
distribution set (not shown) is also below the GRV LO curve by around 20\%.}.
In addition, using $2\ETJ$ for the hard scale instead of $\ETJ/2$ lowers
the cross
section by up to 20\%. When both jets are at equal $\eta$
and equal $E_T^{jet}$,
$\sqrt{\hat{s}} = 2\ETJ$
and so the most common choices of scale are covered by the range
$\ETJ/2$ to $2\ETJ$.
\item
Due to the fact that we are probing low $x$ partons in the proton,
the standard approximation that the incoming partons are collinear and
on-shell may be invalid. Analytic estimates of the effect of allowing the
incoming partons to develop non-zero transverse momentum ($k_T$) have been
made in ~\cite{jeff2} using a so-called `$k_T$ factorization'
prescription. Calculations of the cross section using this technique
are so far available only for the photon-gluon fusion contribution to the
direct photon cross section.
The typical $k_T$ developed is of the order of a GeV.
In Fig.~\ref{f:final2}b
we show again the data of Fig.~\ref{f:final2}a, and the standard LO QCD
calculation using the GRV LO proton parton distributions.
Also shown are the
curves for the same parton distribution set
for the $x_\gamma^{LO}=1$ contribution alone,
and for just the gluon-induced part of the
$x_\gamma^{LO}=1$ cross section.
% but including only the direct
% photon contribution, and the same calculation including only the
% gluon-induced
% part of the direct photon cross section.
(From this it can be seen that
according to LO QCD, most of the direct photon cross section is
attributable to the
photon-gluon fusion diagram.)
This
latter
curve may then be compared to the final
curve, which shows the `$k_T$' result for the same cross section.
Non-zero parton $k_T$ in this prescription lowers the cross section
by as much as 30\%, and could bring a full calculation into better agreement
with the data.
\item
Non-perturbative `hadronisation' effects can be expected to be significant
for jets.
In Fig.~\ref{f:final2}c we show the data of
Fig.~\ref{f:final2}a compared to Monte Carlo estimates of the cross section
calculated from
partons generated with HERWIG using the
GRV LO proton parton distribution set and the
LAC1~\cite{LAC} photon parton
distribution set.
% Also shown is the cross section calculated using the partons
% after the addition of $k_T$ in the Monte Carlo simulation. This $k_T$ comes
% from the initial state parton showering and from the default addition of
% intrinsic
% $k_T$ with a gaussian distribution with width 0.7~GeV.
Also shown is the
cross section
obtained by performing jet finding on the simulated
final state particles. This histogram agrees with the shape
of the
data better than the analytic LO QCD calculations of Fig.~\ref{f:final2}a.
Over most of the $\bar{\eta}$ region these histograms lie within
20\% of each other.
\end{itemize}

Considering the uncertainties in LO calculations arising
from the choice of the hard scale (taken to be $\ETJ/2$ in these
calculations) and the fact
that hadronisation effects are not included in the theoretical curves,
the description of the measured points by the theoretical
curves is reasonable. The effects of $k_T$, hadronisation, and
of using different
parton distributions are of comparable magnitude.
That the direct jet cross section is sensitive to the parton densities of
the proton can be seen by comparing different Monte Carlo estimates of this
curve generated using
different proton parton distribution sets.
In Fig.~\ref{f:final2}d, we show the data compared to hadronic jet
cross sections estimated with HERWIG using the GRV LO and MRSD$_0$ proton
parton distribution sets. The LAC1 photon parton distribution set was
used in both cases for the LO resolved $x_\gamma^{OBS} \ge 0.75$
contribution. The separation between parton distribution sets remains after
hadronisation, and the MRSD$_0$ parton distribution gives a
consistently lower cross section for negative values of $\bar{\eta}$.

In summary, the LO QCD predictions for the direct cross section are
consistent with the data at the level of 30\%.
The theoretical cross section is sensitive to the choice of
proton parton distribution function.
Complete NLO calculations will reduce the ambiguity in comparisons to
the data.

\subsection{Resolved photon dijet cross section}

\begin{figure}
\epsfysize=400pt
\epsfbox[-35 250 465 550]{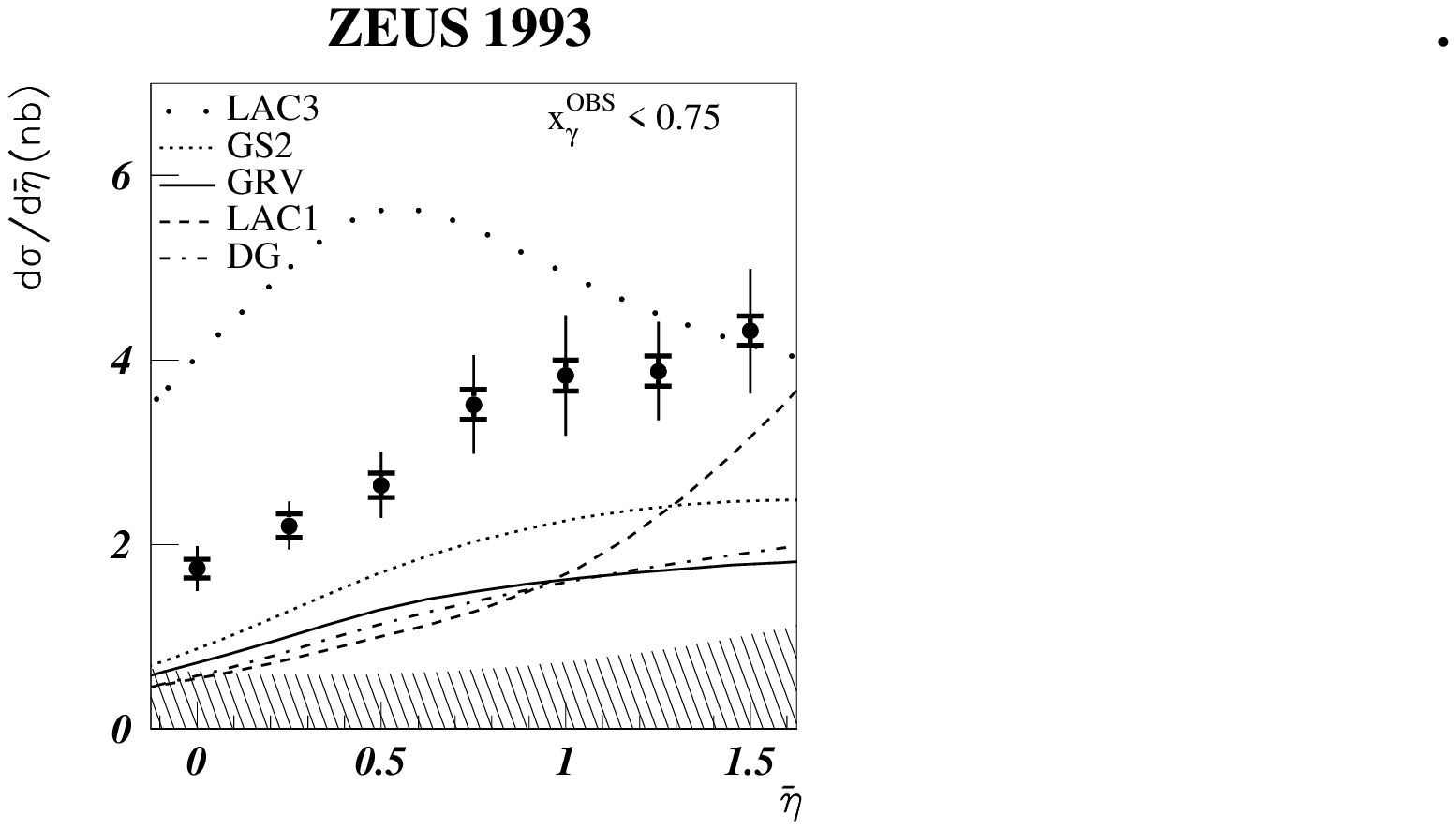}
\caption{\label{f:final3}
            { $d\sigma/d\bar{\eta}$ for  $ep \rightarrow eX + \mbox{2 (or more)
jets},
\DETA < 0.5,
\ETJ > 6$~GeV,
$0.2 < y < 0.8,
Q^2 < 4$~GeV$^2$,
\xgo $< 0.75$.
The solid circles are corrected ZEUS data.
The inner error bars indicate the
statistical errors, the outer error
bars show the systematic uncertainty
(excluding the correlated uncertainty)
added in quadrature.
The shaded band shows the correlated uncertainty from measurement of
energy in the calorimeter and the integrated luminosity.
Also shown are LO QCD calculations.
The parton distribution sets used for the
photon are LAC3, GS2, GRV,
LAC1 and DG. The proton parton distribution
set used
is the MRSA set.}}
\end{figure}

The acceptance correction for the resolved photon process as
defined in Section~2 with a cut of  \xgo$< 0.75$ has been evaluated in the
same way as in the direct photon cross section measurement.
The efficiency and purity are now evaluated for bins with centres in the range
$0.0 < \bar{\eta} < 1.5$.
The efficiency and purity are both approximately flat and around 40\%, giving
an acceptance correction factor of around unity across the whole range of
$\bar{\eta}$.
As for the direct measurement,
the purity depends mostly on migrations across  the $E_T^{jet}$ cut.
No subtraction of jet pedestal energy is carried out.
However, variations in the kinematic selection of the analysis
now give systematic variations in the cross section which are larger
than the statistical errors, and thus the systematic error bars are
correspondingly larger.
% The major systematic effect arises from differences
% between the
% acceptance correction factors caclulated using HERWIG and PYTHIA,
% particularly in the more
% forward region.
As was seen in Fig.~\ref{f:xg}c, both Monte Carlo simulations fail to
describe the
forward region of the
transverse energy flow around the jet axis,
and a reduction of these
systematic
uncertainties
will require improvements in the simulations used.
In Fig.~\ref{f:final3} the correlated uncertainty
arising from the description of the calorimeter
energy
response in the Monte Carlo is again shown separately
by the shaded band. The band also includes the 3.3\% uncertainty in the
measurement of the
integrated luminosity.
The shaded band represents the width of
the uncertainty around each data point.

Table~\ref{t:sys3} and
Fig.~\ref{f:final3} shows the measured cross section
$d\sigma/d\bar{\eta}$ for \xgo $< 0.75$
and the kinematic range indicated in the figure caption.
%\footnote{Numerical
%values for
%these cross sections can be found in a table in the preprint DESY 95-XXX.}.
The cross section is around 2 nb for central values of $\bar{\eta}$ and
rises to
4 nb at $\bar{\eta}=1.5$.
The LO cross sections, shown for
comparison, are calculated using different photon parton distribution sets
and the MRSA set
for the proton. The theoretical sensitivity to the parton distributions in
the proton is small (not shown), with the variations between curves calculated
using different parton distribution sets being much less than the estimated
errors
on the measured cross section.
As \xp\ is higher in these events than in the direct, the effect of $k_T$
from partons in the proton on this calculation is expected to be small.

\begin{table}
\centering
\begin{tabular}{|r|c|c|c|c|}  \hline
$\bar{\eta}$ & $d\sigma/d\bar{\eta}$ & Statistical & Systematic & Energy Scale
Uncertainty \\
             &         (nb)         & Error (nb)  & Error (nb) & (nb)
   \\ \hline
   0.00       &    1.74              & 0.10        & 0.22       & 0.66
    \\
   0.25      &    2.20              & 0.13        & 0.23       & 0.54
   \\
   0.50       &    2.64              & 0.13        & 0.33       & 0.59
    \\
   0.75      &    3.51              & 0.16        & 0.51       & 0.71
   \\
   1.00       &    3.83              & 0.17        & 0.63       & 0.66
    \\
   1.25      &    3.88              & 0.16        & 0.51       & 0.91
   \\
   1.50       &    4.32              & 0.16        & 0.66       & 0.99
    \\  \hline
\end{tabular}
\caption{\label{t:sys3} $d\sigma/d\bar{\eta}$ for $ep \rightarrow eX +
\mbox{2 (or more) jets},
\DETA < 0.5,
\ETJ > 6$~GeV,
$0.2 < y < 0.8$,}
$Q^2 < 4$~GeV$^2$, \xgo $< 0.75$.
\end{table}

The DG~\cite{DG}, GRV LO and GS2 parton distribution sets reproduce the shape
of the cross section well and can be brought into agreement with the data by
applying a multiplicative factor of 1.5 to 2.
%Factors of this size may reasonably
%be expected to come from NLO calculations, some of which are available for
%similar processes
(Note that the available NLO calculations~\cite{NLO} for inclusive
jet photoproduction,
which is dominated by the resolved process, differ from LO calculations by a
factor of up to two.)
The LO calculations using the LAC1 and LAC3 parton
distribution sets cannot be brought into agreement with the data by a constant
normalisation factor. However, since the jet
pedestal energies are not well described by the Monte Carlo
for the resolved process, it
is difficult to estimate what the effects of hadronisation and parton
showering might be on the shapes of the curves, and so no parton
distribution set can be
completely excluded at this stage, although LAC3 is disfavoured.

\section{Conclusions}
Differential dijet cross sections have been measured in
photoproduction with the ZEUS detector at HERA.
The cross sections measured are $d\sigma/d\bar\eta$,
$\DETA < 0.5$, $\ETJ > 6$~GeV
and $0.2 < y < 0.8$, in the regions of \xgo $\ge 0.75$
(direct photoproduction)
and \xgo $<0.75$
(resolved photoproduction).
The measured cross sections have been defined in such a way that they are
calculable to higher orders in QCD.
We have corrected back to the final state particles
and no subtraction of jet pedestal energies has been carried out.
Our results are compared with the expectations of LO QCD.
The direct cross section, which is sensitive to the gluon content of the
proton, is consistent with LO QCD calculations to within 30\%. Hadronisation,
incoming parton $k_T$ and the choice of scale also influence this comparison.
%exhibits a different shape to that predicted by LO QCD calculations using
%parton distribution sets fitted to existing data.
%This effect could be explained by non-zero parton $k_T$ at low $x_p$,
%other higher order effects, or hadronisation.
The shape of the resolved cross section, which is sensitive to the gluon
content of the photon, is described by LO QCD calculations using the
DG, GRV LO and GS2 photon parton distributions. However, most LO QCD
calculations lie below the data by a factor of 1.5 to 2.
Comparison with NLO calculations for both the direct and
resolved photon cross sections would be extremely valuable and
should allow stronger conclusions to be drawn.

\section*{Acknowledgements}
We thank the DESY Directorate for their strong support and
encouragement and the HERA machine group for providing
colliding beams. We acknowledge the assistance of the DESY
computing and networking staff.
It is also a pleasure to thank J. R. Forshaw and
R. G. Roberts for useful discussions and for providing
calculations based upon their publications.

%--------- REFERENCES -------------

\clearpage
\newpage

\clearpage

\clearpage

\clearpage

\end{document}